\documentclass[preprint2,times]{aastex631}

\PassOptionsToPackage{dvipsnames,svgnames,x11names}{xcolor}
\usepackage{tabularx}

\usepackage{amsmath}  
\usepackage{amssymb}  
\usepackage[T1]{fontenc}
\usepackage{xspace}
\usepackage{multirow}

\defcitealias{kenworthy_salt3_2021}{K21}

\usepackage{newunicodechar,graphicx}
\DeclareRobustCommand{\okina}{%
  \raisebox{\dimexpr\fontcharht\font`A-\height}{%
    \scalebox{0.8}{`}%
  }%
}
\newunicodechar{ʻ}{\okina}

\newif{\ifchangetext}
\changetextfalse

\InputIfFileExists{\jobname-options}

\ifchangetext
  
  \newcommand{\changenote}[1]{\textcolor{blue}{ \bf #1}}x
\else
  
  \newcommand{\changenote}[1]{}
\fi

\hypersetup{
    colorlinks=True,
    citecolor=black,
    linkcolor=black,
    filecolor=magenta,      
    urlcolor=cyan,
}

\newcommand{\etal}{{et al.~}}

\def\arcsec{\ensuremath{^{\prime\prime}}}

\def\xone{\ensuremath{x_{1}}}
\def\C{\ensuremath{\mathcal{C}}}

\newcommand{\lensedsn}{SN\,H$0$pe\xspace}

\newcommand{\sntd}{{\fontfamily{qcr}\selectfont{SNTD}}\xspace}

\newcommand{\hst}{\textit{HST}\xspace}

\newcommand{\webb}{\textit{JWST}\xspace}

\newcommand{\snz}{1.783}

\usepackage{xspace}

\newcommand{\dtab}{$-116.6^{+10.8}_{-9.3}$\xspace}
\newcommand{\dtcb}{$-48.6^{+3.6}_{-4.0}$\xspace}
\newcommand{\mua}{$4.3^{+1.6}_{-1.8}$\xspace}
\newcommand{\mub}{$7.6^{+3.6}_{-2.6}$\xspace}
\newcommand{\muc}{$6.4^{+1.6}_{-1.5}$\xspace}

\begin{document}

\title{JWST Photometric Time-Delay and Magnification Measurements for the Triply-Imaged Type Ia ``Supernova H0pe'' at $\mathbf{z=1.78}$}

\author[0000-0002-2361-7201
]{J.~D.~R.~Pierel}
\correspondingauthor{J.~D.~R.~Pierel} 
\email{jpierel@stsci.edu}
\affil{Space Telescope Science Institute, 3700 San Martin Drive, Baltimore, MD 21218, USA}
\affil{NASA Einstein Fellow}

\author[0000-0003-1625-8009
]{B.~L.~Frye} 
\affiliation{Department of Astronomy/Steward Observatory, University of Arizona, 933 N. Cherry Avenue, Tucson, AZ 85721, USA
}

\author[0000-0002-2282-8795]{M.~Pascale} 
\affiliation{Department of Astronomy, University of California, 501 Campbell Hall \#3411, Berkeley, CA 94720, USA
}

\author[0000-0001-6052-3274]{G.~B.~Caminha} 
\affiliation{Technical University of Munich, TUM School of Natural Sciences, Department of Physics, James-Franck-Str.~1, 85748 Garching, Germany}
\affiliation{Max-Planck-Institut f\"ur Astrophysik, Karl-Schwarzschild-Str. 1, D-85748 Garching, Germany 
}

\author[0000-0003-1060-0723]{W.~Chen} 
\affiliation{Minnesota Institute for Astrophysics, 116 Church St SE, Minneapolis, MN 55455
}
\affiliation{Department of Physics,
Oklahoma State University, 145 Physical Sciences Bldg, Stillwater, OK
74078, USA}

\author{S.~Dhawan} 
\affiliation{Institute of Astronomy and Kavli Institute for Cosmology, University of Cambridge, Madingley Road, Cambridge CB3 0HA, UK
}

\author[0000-0002-5116-7287  ]{D.~Gilman} 
\affiliation{Department of Astronomy $\&$ Astrophysics, University of Chicago, Chicago, IL 60637, USA }
\affiliation{Brinson Fellow}

\author[0000-0002-6741-983X]{M.~Grayling} 
\affiliation{Institute of Astronomy and Kavli Institute for Cosmology, University of Cambridge, Madingley Road, Cambridge CB3 0HA, UK
}

\author[0000-0002-6741-983X]{S.~Huber} 
\affiliation{Max-Planck-Institut f\"ur Astrophysik, Karl-Schwarzschild-Str. 1, D-85748 Garching, Germany 
}
\affiliation{Technical University of Munich, TUM School of Natural Sciences, Department of Physics, James-Franck-Str.~1, 85748 Garching, Germany}

\author[0000-0003-3142-997X]{P.~Kelly} 
\affiliation{Minnesota Institute for Astrophysics, 116 Church St SE, Minneapolis, MN 55455
}

\author[0009-0005-6323-0457]{S.~Thorp} 
\affiliation{Oskar Klein Centre, Department of Physics, Stockholm University, SE-106 91 Stockholm, Sweden
}
\author[0000-0001-5409-6480]{N.~Arendse} 
\affiliation{Oskar Klein Centre, Department of Physics, Stockholm University, SE-106 91 Stockholm, Sweden
}

\author[0000-0003-3195-5507]{S.~Birrer} 
\affiliation{Department of Physics and Astronomy, Stony Brook University, Stony Brook, NY 11794, USA
}

\author[0000-0002-1537-6911]{M.~Bronikowski}
\affiliation{Center for Astrophysics and Cosmology, University of Nova Gorica, Vipavska 11c, 5270 Ajdov\v{s}\v{c}ina, Slovenia.
}

\author{R.~Ca\~nameras} 
\affiliation{Max-Planck-Institut f\"ur Astrophysik, Karl-Schwarzschild-Str. 1, D-85748 Garching, Germany 
}
\affiliation{Technical University of Munich, TUM School of Natural Sciences, Department of Physics, James-Franck-Str.~1, 85748 Garching, Germany}

\author[0000-0001-7410-7669]{D.~Coe} 
\affiliation{Space Telescope Science Institute, 3700 San Martin Drive, Baltimore, MD 21218, USA}
\affiliation{Association of Universities for Research in Astronomy (AURA) for the European Space Agency (ESA), STScI, Baltimore, MD 21218, USA}
\affiliation{Center for Astrophysical Sciences, Department of Physics and Astronomy, The Johns Hopkins University, 3400 N Charles St. Baltimore, MD 21218, USA}

\author[0000-0003-3329-1337]{S.~H.~Cohen} 
\affiliation{School of Earth and Space Exploration, Arizona State University,
Tempe, AZ 85287-1404, USA}

\author[0000-0003-1949-7638]{C.~J.~Conselice} 
\affiliation{Jodrell Bank Centre for Astrophysics, Alan Turing Building,
University of Manchester, Oxford Road, Manchester M13 9PL, UK}

\author[0000-0001-9491-7327]{S.~P.~Driver} 
\affiliation{International Centre for Radio Astronomy Research (ICRAR) and the
International Space Centre (ISC), The University of Western Australia, M468,
35 Stirling Highway, Crawley, WA 6009, Australia}

\author[0000-0002-9816-1931]{J.~C.~J.~D\'Silva}
\affiliation{International Centre for Radio Astronomy Research (ICRAR) and the
International Space Centre (ISC), The University of Western Australia, M468,
35 Stirling Highway, Crawley, WA 6009, Australia}
\affiliation{ARC Centre of Excellence for All Sky Astrophysics in 3 Dimensions
(ASTRO 3D), Australia}

\author[0000-0003-0209-674X]{M.~Engesser} 
\affiliation{Space Telescope Science Institute, 3700 San Martin Drive, Baltimore, MD 21218, USA}

\author[0000-0002-7460-8460]{N.~Foo} 
\affiliation{School of Earth and Space Exploration, Arizona State University,
Tempe, AZ 85287-1404, USA}

\author[0000-0002-8526-3963]{C.~Gall} 
\affiliation{DARK, Niels Bohr Institute, University of Copenhagen, Jagtvej 128, 2200 Copenhagen, Denmark
}

\author[0000-0003-3418-2482]{N.~Garuda} 
\affiliation{Department of Astronomy/Steward Observatory, University of Arizona, 933 N. Cherry Avenue, Tucson, AZ 85721, USA
}

\author[0000-0002-5926-7143]{C.~Grillo} 
\affiliation{Dipartimento di Fisica, Universit\`a  degli Studi di Milano, via Celoria 16, I-20133 Milano, Italy}
\affiliation{INAF - IASF Milano, via A. Corti 12, I-20133 Milano, Italy
}

\author[0000-0001-9440-8872]{N.~A.~Grogin} 
\affiliation{Space Telescope Science Institute, 3700 San Martin Drive, Baltimore, MD 21218, USA}

\author{J.~Henderson} 
\affiliation{Institute of Astronomy and Kavli Institute for Cosmology, University of Cambridge, Madingley Road, Cambridge CB3 0HA, UK
}

\author[0000-0002-4571-2306]{J.~Hjorth} 
\affiliation{DARK, Niels Bohr Institute, University of Copenhagen, Jagtvej 128, 2200 Copenhagen, Denmark
}

\author[0000-0003-1268-5230]{R.~A.~Jansen} 
\affiliation{School of Earth and Space Exploration, Arizona State University,
Tempe, AZ 85287-1404, USA}

\author[0000-0001-5975-290X]{J.~Johansson} 
\affiliation{Oskar Klein Centre, Department of Physics, Stockholm University, SE-106 91 Stockholm, Sweden
}

\author[0000-0001-9394-6732]{P.~S.~Kamieneski} 
\affiliation{School of Earth and Space Exploration, Arizona State University,
Tempe, AZ 85287-1404, USA}

\author[0000-0002-6610-2048]{A.~M.~Koekemoer} 
\affiliation{Space Telescope Science Institute, 3700 San Martin Drive, Baltimore, MD 21218, USA}

\author[0000-0003-2037-4619]{C.~Larison} 
\affiliation{Department of Physics and Astronomy, Rutgers University, 136 Frelinghuysen Road, Piscataway, NJ 08854, USA
}

\author[0000-0001-6434-7845]{M.~A.Marshall} 
\affiliation{National Research Council of Canada, Herzberg Astronomy \&
Astrophysics Research Centre, 5071 West Saanich Road, Victoria, BC V9E 2E7,
Canada}
\affiliation{ARC Centre of Excellence for All Sky Astrophysics in 3 Dimensions
(ASTRO 3D), Australia}

\author[0000-0003-3030-2360]{L.~A.~Moustakas} 
\affiliation{Jet Propulsion Laboratory, California Institute of Technology, 4800 Oak Grove Dr, Pasadena, CA 91109
}

\author[0000-0001-6342-9662]{M.~Nonino} 
\affiliation{INAF-Osservatorio Astronomico di Trieste, Via Bazzoni 2, 34124
Trieste, Italy} 

\author[0000-0002-6150-833X]{R.~{Ortiz~III}} 
\affiliation{School of Earth and Space Exploration, Arizona State University,
Tempe, AZ 85287-1404, USA}

\author[0000-0003-4743-1679]{T.~Petrushevska}
\affiliation{Center for Astrophysics and Cosmology, University of Nova Gorica, Vipavska 11c, 5270 Ajdov\v{s}\v{c}ina, Slovenia.
}

\author[0000-0003-3382-5941]{N.~Pirzkal} 
\affiliation{Space Telescope Science Institute, 3700 San Martin Drive, Baltimore, MD 21218, USA}

\author[0000-0003-0429-3579]{A.~Robotham} 
\affiliation{International Centre for Radio Astronomy Research (ICRAR) and the
International Space Centre (ISC), The University of Western Australia, M468,
35 Stirling Highway, Crawley, WA 6009, Australia}

\author[0000-0003-0894-1588]{R.~E.~Ryan, Jr.} 
\affiliation{Space Telescope Science Institute, 3700 San Martin Drive, Baltimore, MD 21218, USA}

\author[0000-0003-2497-6334]{S.~Schuldt} 
\affiliation{Dipartimento di Fisica, Universit\`a  degli Studi di Milano, via Celoria 16, I-20133 Milano, Italy}
\affiliation{INAF - IASF Milano, via A. Corti 12, I-20133 Milano, Italy
}

\author[0000-0002-7756-4440]{L.~G.~Strolger} 
\affiliation{Space Telescope Science Institute, 3700 San Martin Drive, Baltimore, MD 21218, USA}

\author[0000-0002-7265-7920]{J.~Summers} 
\affiliation{School of Earth and Space Exploration, Arizona State University,
Tempe, AZ 85287-1404, USA}

\author[0000-0001-5568-6052]{S.~H.~Suyu
} 
\affiliation{Technical University of Munich, TUM School of Natural Sciences, Department of Physics, James-Franck-Str.~1, 85748 Garching, Germany}
\affiliation{Max-Planck-Institut f\"ur Astrophysik, Karl-Schwarzschild-Str. 1, D-85748 Garching, Germany 
}

\affiliation{Academia Sinica Institute of Astronomy and Astrophysics (ASIAA), 11F of ASMAB, No.1, Section 4, Roosevelt Road, Taipei 10617, Taiwan \label{asiaa}
}

\author[0000-0002-8460-0390]{T.~Treu} 
\affiliation{Physics and Astronomy Department, University of California, Los Angeles CA 90095
}

\author[0000-0001-9262-9997]{C.~N.~A.~Willmer} 
\affiliation{Steward Observatory, University of Arizona,
933 N Cherry Ave, Tucson, AZ, 85721-0009, USA}

\author[0000-0001-8156-6281]{R.~A.~Windhorst} 
\affiliation{School of Earth and Space Exploration, Arizona State University,
Tempe, AZ 85287-1404, USA}

\author[0000-0001-7592-7714]{H.~Yan} 
\affiliation{Department of Physics and Astronomy, University of Missouri,
Columbia, MO 65211, USA}

\author[0000-0002-0350-4488]{A.~Zitrin} 
\affiliation{Department of Physics, Ben-Gurion University of the Negev, P.O. Box 653, Beer-Sheva, 84105, Israel}

\author[0000-0003-3108-9039]{A.~Acebron} 
\affiliation{Dipartimento di Fisica, Universit\`a  degli Studi di Milano, via Celoria 16, I-20133 Milano, Italy}
\affiliation{INAF - IASF Milano, via A. Corti 12, I-20133 Milano, Italy
}

\author[0000-0001-6711-8140]{S.~Chakrabarti} 
\affiliation{Department of Physics and Astronomy, University of Alabama, 301 Sparkman Drive Huntsville, AL 35899
}

\author[0000-0003-4263-2228]{D.~A.~Coulter} 
\affiliation{Space Telescope Science Institute, 3700 San Martin Drive, Baltimore, MD 21218, USA}

\author[0000-0003-2238-1572]{O.~D.~Fox} 
\affiliation{Space Telescope Science Institute, 3700 San Martin Drive, Baltimore, MD 21218, USA}

\author[0000-0001-8156-0330]{X.~Huang} 
\affiliation{Department of Physics \& Astronomy, University of San Francisco, San Francisco, CA 94117-1080
}

\author[0000-0001-8738-6011]{S.~W.~Jha} 
\affiliation{Department of Physics and Astronomy, Rutgers University, 136 Frelinghuysen Road, Piscataway, NJ 08854, USA
}

\author{G.~Li} 
\affiliation{Purple Mountain Observatory, No. 10 Yuanhua Road, Nanjing 210023, People's Republic of China
}

\author{P.~A.~Mazzali} 
\affiliation{Astrophysices Research Institute, Liverpool John Moores University, Liverpool, UK}  \affiliation{Max-Planck-Institut f\"ur Astrophysik, Karl-Schwarzschild-Str. 1, D-85748 Garching, Germany 
}

\author[0000-0002-7876-4321]{A.~K.~Meena} 
\affiliation{Department of Physics, Ben-Gurion University of the Negev, P.O. Box 653, Beer-Sheva, 84105, Israel}

\author[0000-0002-2807-6459]{I.~P\'erez-Fournon} 
\affiliation{Instituto de Astrof\'{\i}sica de Canarias, V\'{\i}a L\'actea, 38205 La Laguna, Tenerife, Spain.} 
\affiliation{Universidad de La Laguna, Departamento de Astrof\'{\i}sica,  38206 La Laguna, Tenerife, Spain.
}

\author[0000-0002-5391-5568]{F.~Poidevin} 
\affiliation{Instituto de Astrof\'{\i}sica de Canarias, V\'{\i}a L\'actea, 38205 La Laguna, Tenerife, Spain.}
\affiliation{Universidad de La Laguna, Departamento de Astrof\'{\i}sica,  38206 La Laguna, Tenerife, Spain.
}
\author[0000-0002-4410-5387]{A.~Rest} 
\affiliation{Space Telescope Science Institute, 3700 San Martin Drive, Baltimore, MD 21218, USA}

\author[0000-0002-6124-1196]{A.~G.~Riess} 
\affiliation{Space Telescope Science Institute, 3700 San Martin Drive, Baltimore, MD 21218, USA}
\affiliation{Center for Astrophysical Sciences, Department of Physics and Astronomy, The Johns Hopkins University, 3400 N Charles St. Baltimore, MD 21218, USA}

\begin{abstract}
Supernova (SN) H0pe is a gravitationally lensed, triply-imaged, Type Ia SN (SN\,Ia) discovered in \textit{James Webb Space Telescope} imaging of the PLCK G165.7+67.0 cluster of galaxies. Well-observed multiply-imaged SNe provide a rare opportunity to constrain the Hubble constant ($H_0$), by measuring the relative time delay between the images and modeling the foreground mass distribution. \lensedsn is located at $z=\snz$, and is the first SN\,Ia with sufficient light curve sampling and long enough time delays for an $H_0$ inference. Here we present photometric time-delay measurements and SN properties of \lensedsn. Using \textit{JWST}/NIRCam photometry we measure time delays of $\Delta t_{ab}=$\dtab and $\Delta t_{cb}=$\dtcb  observer-frame days relative to the last image to arrive (image 2b; all uncertainties are $1\sigma$), which corresponds to a $\sim5.6\%$ uncertainty contribution for $H_0$ assuming $70 \rm{km s^{-1} Mpc^{-1}}$. We also constrain the absolute magnification of each image to $\mu_{a}=$\mua, $\mu_{b}=$\mub, $\mu_{c}=$\muc by comparing the observed peak near-IR magnitude of \lensedsn to the non-lensed population of SNe\,Ia.

\end{abstract}

\section{Introduction}
\label{sec:intro}
Strong gravitational lensing can cause multiple images of a background source to appear, as light propagating along different paths are focused by a foreground galaxy or galaxy cluster \citep[e.g.,][]{narayan_lectures_1997}. Such a phenomenon requires chance alignment between the observer, the background source, and the foreground galaxy. If the multiply-imaged source has variable brightness then depending on the relative geometrical and gravitational potential differences of each path, the source images will appear delayed by hours to weeks (for galaxy-scale lenses) or months to years (for cluster-scale lenses) \citep[e.g.,][]{oguri_strong_2019}. 

Precise measurements of this ``time delay'' yield a direct distance measurement to the lens system that constrains the Hubble constant ($H_0$) in a single step \citep[e.g.,][]{refsdal_possibility_1964,linder_lensing_2011,paraficz_gravitational_2009,treu_time_2016,grillo_measuring_2018,grillo_accuracy_2020,birrer_time-delay_2022,treu_strong_2022,suyu_strong_2023}. Quasars have historically been used for this purpose \citep[e.g.,][]{vuissoz_cosmograil_2008,suyu_dissecting_2010,tewes_cosmograil_2013,bonvin_h0licow_2017,birrer_h0licow_2019,bonvin_cosmograil_2018,bonvin_cosmograil_2019,wong_h0licow_2020}, since SNe exploding in multiply-imaged galaxies are relatively rare with only a handful discovered thus far \citep{kelly_multiple_2015,goobar_iptf16geu_2017,rodney_gravitationally_2021,chen_jwst-ers_2022,kelly_strongly_2022,goobar_sn_2022,pierel_lenswatch_2023}. Nevertheless, SNe have several advantages over quasars that are attractive for cosmological measurements: 1) SNe fade, enabling predictive experiments on the delayed appearance of trailing images and more accurate photometry, as the SN (or quasar), lens, and host fluxes are otherwise highly blended \citep{ding_improved_2021}, 2) SNe generally have predictable behavior compared to quasars, simplifying time-delay measurements and requiring much shorter observing campaigns, and 3) the impact of a ``microlensing time-delay'' is significantly mitigated due to the more compact source size \citep{tie_microlensing_2018,bonvin_impact_2019}.

SNe of Type Ia (SNe\,Ia), known as ``standardizable candles'', can be used to measure cosmological parameters by way of luminosity distances and the cosmic distance ladder \citep[e.g.,][]{garnavich_supernova_1998,riess_observational_1998,perlmutter_measurements_1999,scolnic_complete_2018,brout_pantheon_2022}. SNe\,Ia are of particular value when strongly lensed, as their standardizable absolute brightness can provide additional leverage for lens modeling by limiting the uncertainty caused by the mass-sheet degeneracy \citep{falco_model_1985,kolatt_gravitational_1998,holz_seeing_2001,oguri_gravitational_2003,patel_three_2014,nordin_lensed_2014,rodney_illuminating_2015,xu_lens_2016,foxley-marrable_impact_2018,birrer_hubble_2022}, though only in cases where millilensing and microlensing are not extreme \citep[see][]{goobar_iptf16geu_2017,yahalomi_quadruply_2017,foxley-marrable_impact_2018,dhawan_magnification_2019}. Additionally, SNe\,Ia have well-understood models of light curve evolution \citep{hsiao_k_2007,guy_supernova_2010,saunders_snemo_2018,leget_sugar_2020,kenworthy_salt3_2021,mandel_hierarchical_2022,pierel_salt3-nir_2022} that enable precise time-delay measurements using color curves, which removes the effects of macro- and achromatic microlensing \citep[e.g.,][]{pierel_turning_2019,huber_holismokes_2021,rodney_gravitationally_2021,pierel_lenswatch_2023}.  The first two spectroscopically confirmed multiply-imaged SNe\,Ia were iPTF16geu \citep{goobar_iptf16geu_2017} and SN Zwicky \citep{goobar_sn_2022,pierel_lenswatch_2023}, but both had very short time-delays of $\sim0.25$-$1.5$ days that precluded competitive $H_0$ measurements \citep{dhawan_magnification_2019,pierel_lenswatch_2023}. 

Overall, while the advantages of using SNe for time-delay cosmography relative to other probes have been well-documented \citep[e.g.,][]{refsdal_possibility_1964,kelly_multiple_2015,goobar_iptf16geu_2017, goldstein_precise_2018,huber_strongly_2019,pierel_turning_2019,suyu_holismokes_2020,pierel_projected_2021,rodney_gravitationally_2021,suyu_strong_2023}, only a single strongly lensed SN \citep[the core-collapse SN ``Refsdal'',][]{kelly_multiple_2015} has had time delays long enough ($\sim$1\,year) and sufficient light curve sampling for precise time-delay measurements ($\sim$1.5\%; \citealt{kelly_magnificent_2023}), which yielded a $\sim$6\% measurement of $H_0$ (64.8$_{-4.3}^{+4.4}$ or 66.6$_{-3.3}^{+4.1 }$ km s$^{-1}$ Mpc$^{-1}$, depending on lens model weights; \citealt{kelly_constraints_2023}). 

\begin{figure*}
    \centering
        \includegraphics[trim={0cm 0cm 0cm 0cm},clip,width=\linewidth]{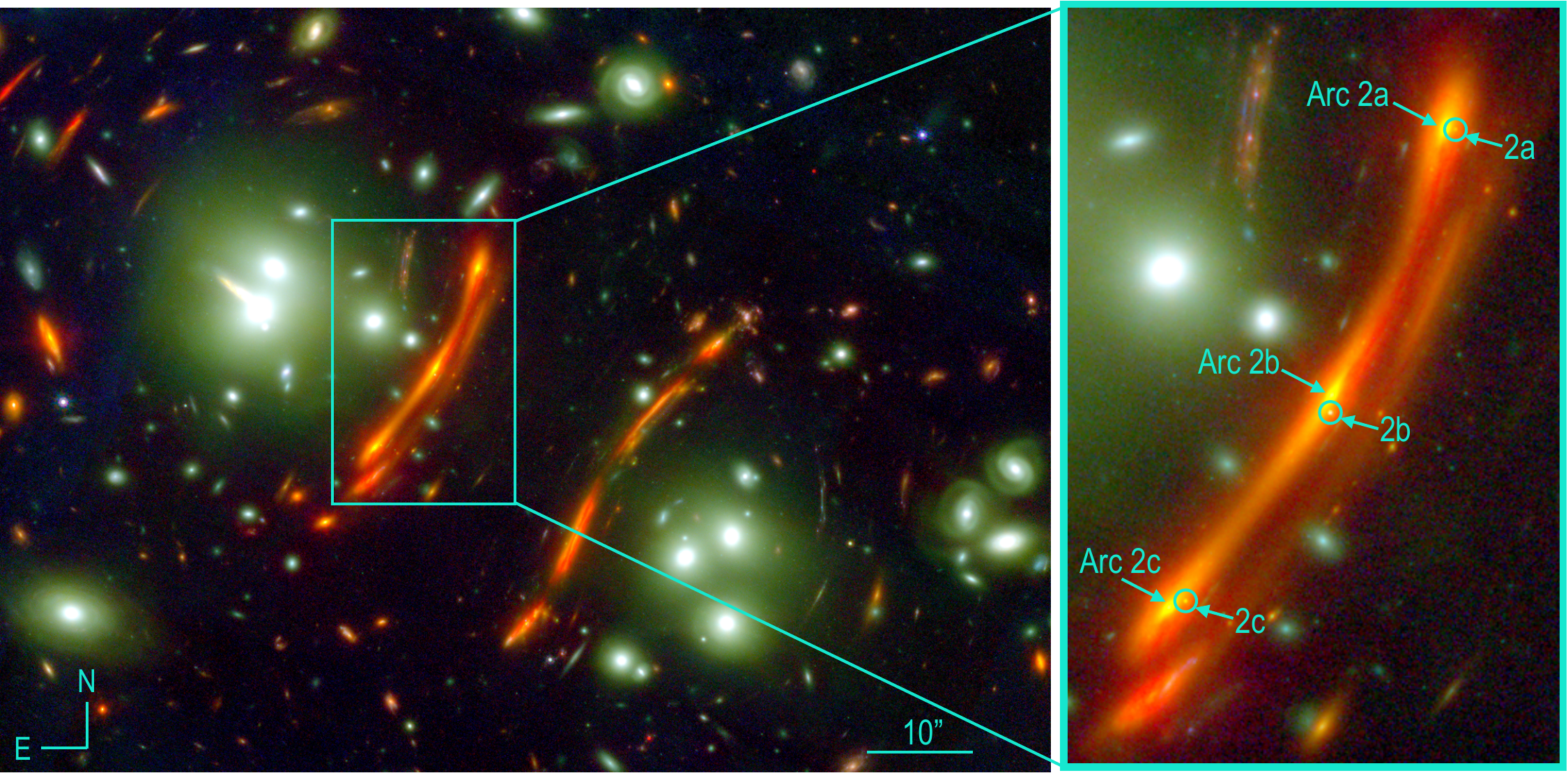}
    \caption{ {\it Left}: {\it JWST}/NIRCam color image in the central region of PLCK G165.7+67.0 from PEARLS (epoch 1).  The filters used are F090W (blue), F200W (green), and F444W (red). The images were drizzled to $0\farcs03/\rm{pix}$, and the image scale and orientation are as shown.  {\it Right:} Closeup of the boxed region depicting the three images of the galaxy Arc~2, as labeled.  \lensedsn (circled) appears in all three images and flips parity between images 2c and 2b and between images 2b and 2a, as predicted by lensing theory.}
    \label{fig:color_im}
\end{figure*}

A new multiply-imaged SN\,Ia, \lensedsn \citep[][hereafter F24]{frye_jwst_2024}, was discovered in 2023 March by the \textit{James Webb Space Telescope} (\textit{JWST}) program, ``Prime Extragalactic Areas for Reionization and Lensing Science'' \citep[PEARLS; PID 1176,][]{windhorst_jwst_2023}, in NIRCam observations of the PLCK G165.7+67.0 cluster of galaxies (Figure \ref{fig:color_im}).  Previous observations of the PLCK G165.7+67.0 cluster are described in \citet{canameras_plancks_2015, canameras_plancks_2018,harrington_early_2016,frye_plck_2019,pascale_possible_2022}. This work is part of a series of papers analyzing the \lensedsn system, which together with \citet{pascale_h0_2024} will provide the first measurement of $H_0$ with a multiply-imaged SN\,Ia. The discovery of \lensedsn is described in F24, and time delays derived only from spectroscopy of the system are presented in Chen et al. (2024). In this paper we focus on imaging of the SN itself, including the photometric time-delay measurement critical to that forthcoming $H_0$ inference. These measurements are made using NIRCam photometry and well-vetted model for SN\,Ia light curve evolution, and we produce a series of simulations to ensure the uncertainties are well-characterized. We also compare the observed brightness of \lensedsn to the general SN\,Ia population to infer absolute magnifications for the system. 

The outline of this paper is as follows: Section \ref{sec:obs} gives an overview of \lensedsn and summarizes the \webb observation characteristics, which are described in more detail in F24. Section \ref{sec:photometry} details the photometric measurements used in the remainder of the paper, with analysis of time delays and magnifications reported in Section \ref{sec:fitting}. In Section \ref{sec:sims} we create a suite of simulations to understand various systematic uncertainties associated with the results of Section \ref{sec:fitting}, and we conclude with a discussion of the implications for these measurements in Section \ref{sec:conclusion}. 

\section{Summary of \textit{JWST} Observations}
\label{sec:obs}

The description of PEARLS and Space Telescope Science Institute's Director's Discretionary Time (DDT) observations, SN discovery, and initial analysis of the \lensedsn lensing system are presented by F24. Briefly, the SN was discovered by the PEARLS team in (eight-filter) NIRCam imaging taken 2023 March 30 (MJD $60033$) by comparing the F150W data with an archival \hst WFC3/IR F160W image \citep[2016 May 30, MJD $57538$][]{frye_sn_2023}. These two filters are extremely well-matched in wavelength and transmission, and as the source was very bright ($\lesssim24$\,mag~AB) in F150W but absent in F160W its transient nature was confirmed. The initial photometric estimate of the host galaxy redshift was $z\sim2.3$ from \hst and \textit{Spitzer} $3.6\mu$m$/4.5\mu$m photometry \citep{pascale_possible_2022}, but the poorly-matched resolution between \textit{HST} and \textit{Spitzer} led to this being an overestimate. Updated NIRCam photometry provided a much more robust estimate of $z\sim1.8$, benefiting both from the improved wavelength range and resolution. Adopting predicted time delays and magnifications from the \citet{frye_plck_2019} lens model updated with the NIRCam imaging, initial photometric classification suggested the SN was likely of Type Ia. A \webb DDT proposal was subsequently submitted and accepted (PID 4446), which provided NIRSpec Prism, G140M, and G235M spectroscopy of the multiple images and two epochs of additional NIRCam imaging in six filters (see Section \ref{sub:final_phot} for exposure times). 

\begin{figure}
    \centering
    \includegraphics[width=\columnwidth]{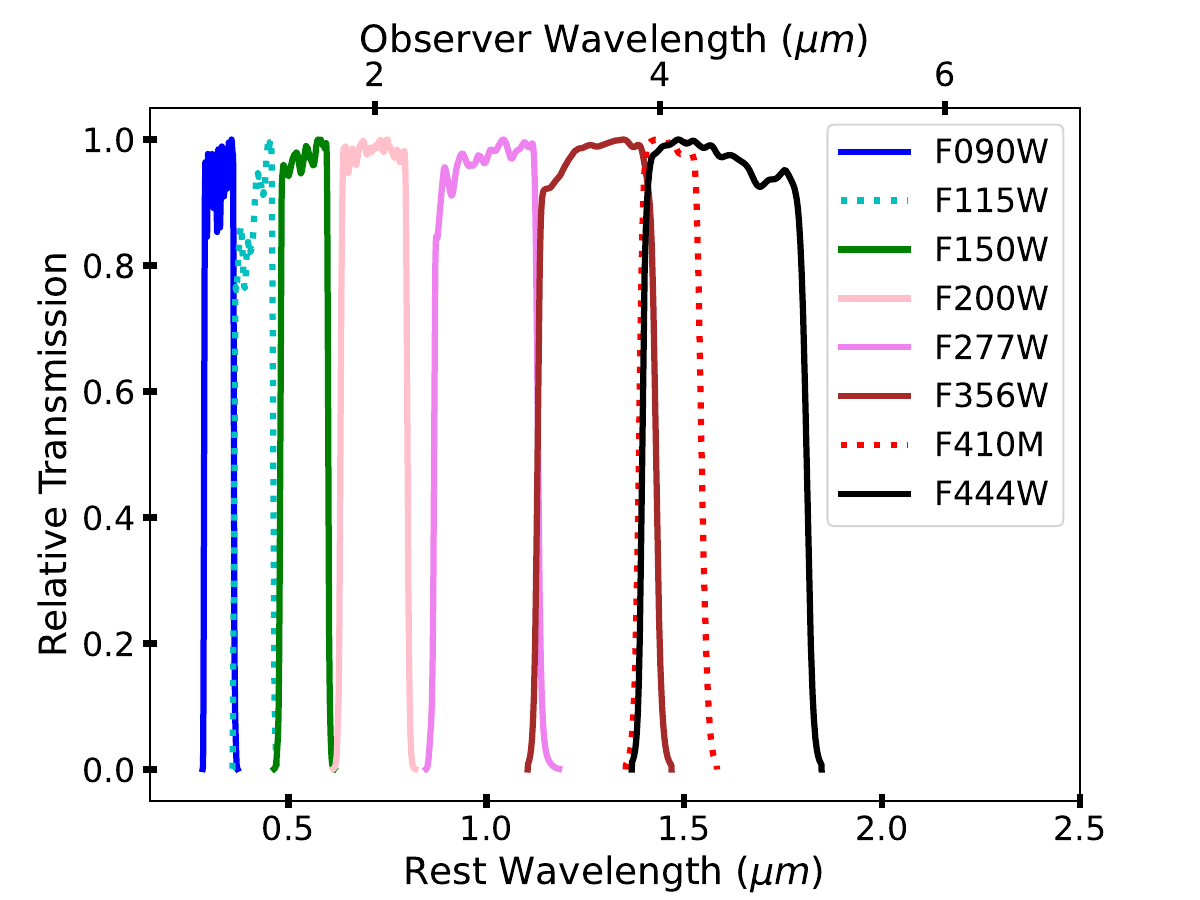}
    \caption{The \webb filters used to observe \lensedsn, with rest-frame wavelength on the lower axis and observer-frame wavelength of the upper axis. Filters with dashed lines (F115W and F410M) were only observed in epoch 1 while the others were observed in all three epochs, but overall rest-frame wavelength coverage is excellent.}
    \label{fig:filters}
\end{figure}

The spectroscopic analysis and updated lens modeling are described in companion papers, but the NIRSpec data unambiguously confirmed the SN was of Type Ia at a refined spectroscopic redshift of $z=\snz$ (Chen et al., 2024 and F24), whose value matches the redshift of the SN host galaxy of $z=1.7833\pm0.0005$ \citep[F24][]{polletta_spectroscopy_2023}. The two NIRCam imaging epochs occurred on 2023 April 22 (MJD $60056$) and 2023 May 9 (MJD $60073$). This meant the light curve for each image contained three observations each separated by $\sim3$ observer-frame weeks, or $\sim1$ rest-frame week. The combined dataset effectively produces a SN\,Ia light curve with nine epochs and rest-frame UV-NIR wavelength coverage, which is sufficient to accurately measure phases of each image and therefore the relative time delays \citep[e.g.,][]{pierel_turning_2019}. The wavelength coverage of the NIRCam imaging is shown in Figure~\ref{fig:filters}.

\section{Measuring Photometry}
\label{sec:photometry}

In most analyses involving dithered exposures, the individual exposures for each filter are drizzled together to create a single (level 3) image. When one is primarily concerned with precisely measuring the position and brightness of a SN, point-spread function (PSF) fitting on the (level 2) NIRCam ``CAL'' images is the best understood method \citep[e.g.,][]{rigby_science_2023}, which preserves the PSF structure compared to drizzled images. CALs are individual exposures that have been calibrated using the STScI JWST Pipeline\footnote{\url{https://github.com/spacetelescope/jwst}}
\citep{bushouse_jwst_2022},  
%
%
and have been bias-subtracted, dark-subtracted, and flat-fielded but not yet corrected for geometric distortion. The NIRCam pixel area map (PAM) for the corresponding short-wavelength (SW) and long-wavelength (LW) modules containing the SN must also be applied to each exposure to correct for pixel area variations across the images. 

Due to the increased brightness of the host galaxy Arc 2 and diminished  NIRCam resolution at longer wavelengths, the S/N of \lensedsn is only high enough to rely on individual exposures in the SW channel, and most reliably in the F150W filter where \lensedsn is brightest. We will therefore measure photometry in the F150W filter on CAL images as a baseline, since this well-established method should be robust (see Section \ref{sub:psf_cal}), but in order to be consistent across all filters this is only used for comparison and not in the final analysis. Instead, we implement a PSF fitting routine suitable for drizzled images in Section \ref{sub:psf_drz}, where we test that the results are consistent with the CAL measurements. Finally, to enable accurate photometry at all wavelengths without a template image, we remove Arc 2 light in the drizzled images using a host surface brightness modeling technique described in Section \ref{sub:psf_host_sub}. We use injected PSF models to ensure the host-subtraction method does not introduce biases, and extend the method to all filters for our final photomery in Section \ref{sub:final_phot}. In all cases we use the PSF models provided by WebbPSF (version 1.2.1)\footnote{\url{https://www.stsci.edu/jwst/science-planning/proposal-planning-toolbox/psf-simulation-tool}} to represent the (level 2) PSF, which have been updated to better match the observed PSF in each filter, and take into account temporal and spatial variation across the detector. 


\subsection{PSF Fitting on F150W CAL Images}
\label{sub:psf_cal}
For each epoch in the F150W filter, we use the Python package \texttt{space\_phot}\footnote{\url{space-phot.readthedocs.io}} to simultaneously constrain the (common) SN flux in all CALs (four, one for each dither position) for all three SN images. Each PSF was fit to the multiple SN images within a $3\times3$ pixel square in an attempt to limit the contamination of the Arc 2 halo. The final measured flux is the integral of each full fitted PSF model, which includes a correction to the infinite aperture flux. These total fluxes, which are in units of MJy/sr, are converted to AB magnitudes using the native pixel scale of each image ($0.03\arcsec/\rm{pix}$ for SW, $0.06\arcsec/\rm{pix}$ for LW). A constant background is removed from the fitting region, which is small enough that a constant background is sufficient. The background is estimated as the median value in an annulus centered at the SN position with an inner radius of 5 pixels and outer radius 7 pixels. In this case the method of background estimation is not critical, as long as it is reasonable and the same method is used in the following section, as the final photometry uses the host galaxy surface brightness modeling method (Section \ref{sub:psf_host_sub}). The measured photometry from this section is only used as a baseline to compare with the F150W photometry measured on drizzled images in the next section, and is not used in the final analysis.

\subsection{PSF Fitting on F150W Drizzled Images}
\label{sub:psf_drz}
In this section we use \texttt{space\_phot} to measure PSF photometry on level 3 F150W drizzled images as a test of our method.  \texttt{space\_phot} creates files with the same header and WCS information as the individual CAL files, but containing only a PSF model at the SN location. The PSF/CAL files are then drizzled using the same method used to produce the drizzled data files, which builds a PSF model suitable for measuring photometry only in that single drizzled image and at the SN position. We then fit the SN flux in the same $3\times3$ square with the same background estimation method as the previous section, and convert to AB magnitudes in the same manner. The resulting magnitudes agree with the F150W CAL file measurements from the previous section to within $\sigma_{\rm{PSF}}=0.04$\,mag. This is within the statistical uncertainty of our level 3 PSF fitting for most filters, but we include it as a possible systematic in our final photometry and proceed with measuring photometry on level 3 drizzled images for the remainder of this work. 

\subsection{Host Model-Subtracted Images}
\label{sub:psf_host_sub}

\begin{figure}
\includegraphics[trim={0cm 0cm 0cm 0cm},clip,width=\linewidth]{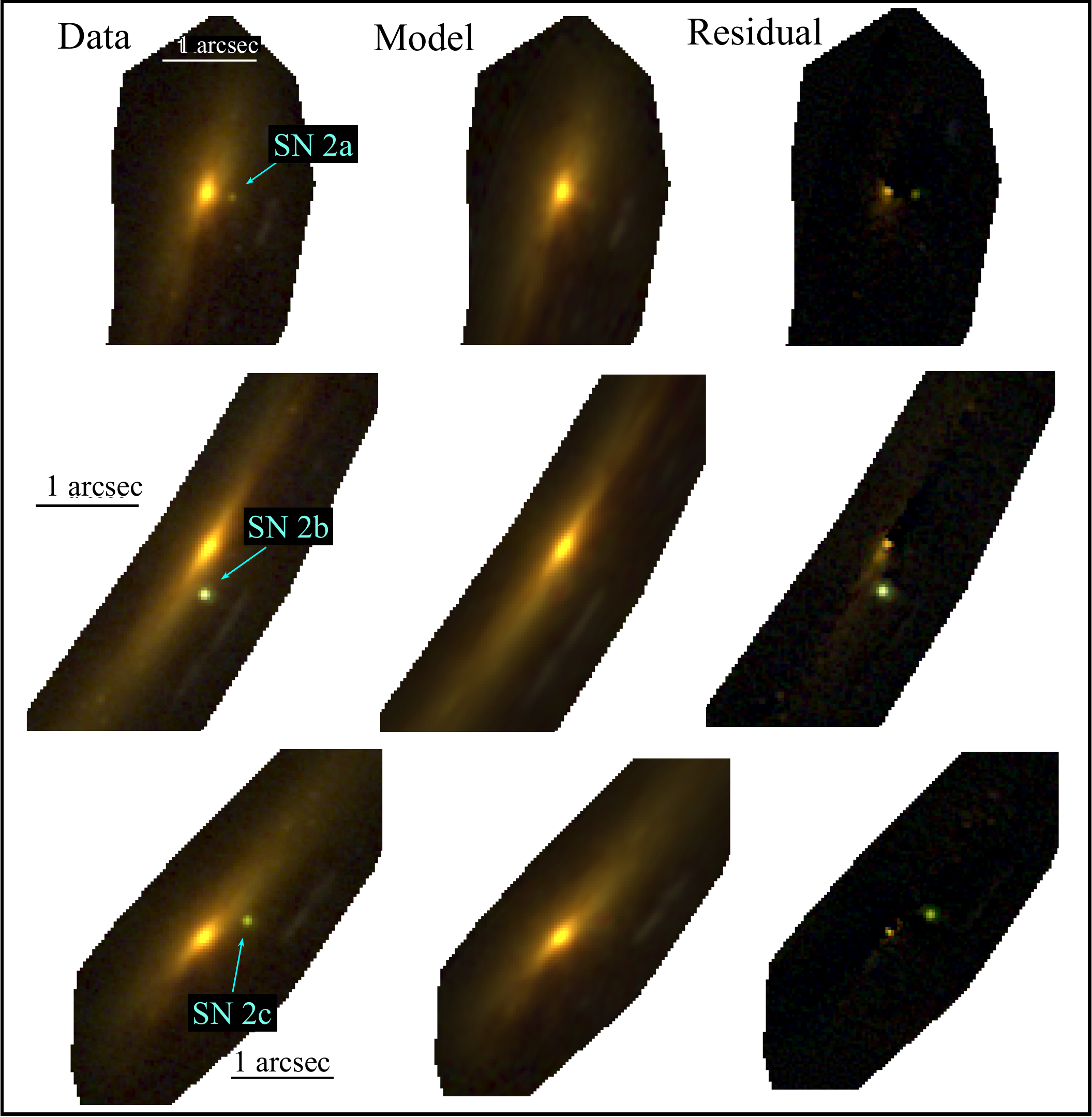}
\caption{Host light subtraction for Arc 2. Color composite images of epoch 1 using the filters F090W, F150W and F200W as blue, green and red, respectively. Left panels show the original data, middle panels show the surface brightness distribution predicted by our strong lensing total mass model, and right panels show the residuals (i.e. data$-$model) with the same scaling. Note that the SN is not present in the model, and so it is expected to be present in the residuals. The orientation is the same as in Figure~\ref{fig:color_im}.}
    \label{fig:host_sub}
\end{figure}

As shown in Figure \ref{fig:color_im}, Arc 2 is bright ($m_{F200W}=20.3$) and red. This makes extraction of the SN light difficult without a template used for difference imaging, particularly at longer wavelengths.
To achieve an efficient subtraction of the contaminant light, we need to perform a surface brightness reconstruction of the host/background galaxy accounting for the lensing effect. Such additional distortions of the arc image are not well represented by analytical profiles, and we therefore use the Gravitational Lens Efficient Explorer (GLEE) software because of its capability to perform detailed surface brightness reconstruction in lensing scenarios \citep{suyu_halos_2010,suyu_disentangling_2012}. The detailed galaxy surface brightness and lens modelling will be published in Caminha \etal, (in preparation) and here we give a short description of the main steps.
We first use the positions of substructures along Arc~2 and other multiply-imaged galaxies to model the gravitational lensing deflection adopting a parametric approach for the cluster total mass distribution. Our total mass parameterization consists of two main cluster scale components plus a dual pseudo-isothermal elliptical (dPIE) density profile for each securely identified cluster member using the F150W magnitudes for the scaling relation, which is similar to previous works \citep[see e.g.][]{pascale_possible_2022}.
In a second step we use the extended light of Arc 2 to model the cluster lens properties and the source light simultaneously, using the best-fit mass model obtained with the multiple image positions as a starting point to find the final solution.
The light distribution of Arc 2 in the source plane is described by a grid of pixels but with a curvature regularization to avoid high pixel-to-pixel variations \citep[see][for more details]{suyu_bayesian_2006}.
The light distribution is then mapped to the image plane and convolved with the PSF estimated directly on each single band image interactively to minimise the differences between the observed data and model.
In Figure \ref{fig:host_sub}, we show an example of subtraction using epoch 1 data. The process is repeated for all epochs and all filters, which are used for the final photometry measured in Section \ref{sub:final_phot}.

    
    
    
    
    

\subsubsection{Injected Source Recovery Test}
\label{sub:fakes}
In order to test if the host subtraction method above biases our SN photometry measurements, we plant three point sources representing SNe around each image of Arc 2 for all epochs and filters. Each injected point source is roughly the same brightness as the corresponding image of \lensedsn, and together the three fake sources and \lensedsn form a rectangle around each Arc 2 image. The locations were chosen to obtain similar contamination from the host galaxy in image 2a, and then those plant positions were translated to positions in 2b and 2c with the lens model. After subtracting the host light derived in the previous section, we measure the flux of each injected PSF with the same method described above. We use the resulting RMS in each filter as an additional systematic uncertainty on our final measured photometry (Table \ref{tab:plant_sigmas}), described in the following section. Note that F356W and F444W are absent, as we found the source reconstruction method was not reliable at these longest wavelengths when recovering injected PSF models. We therefore remove these filters from the current analysis until a template is obtained for the system in \textit{JWST} Cycle 3 (PID 4744) for traditional difference imaging.

\begin{table}
    \centering
    \caption{\label{tab:plant_sigmas} Systematic photometry uncertainties measured by recovering injected PSF models.}
    
    \begin{tabular*}{\linewidth}{@{\extracolsep{\stretch{1}}}*{2}{c}}
\toprule
Filter&\multicolumn{1}{c}{$\sigma_{\rm{plant}}$ (mag)}\\
\hline
F090W&0.02\\
F115W&0.05\\
F150W&0.06\\
F200W&0.07\\
F277W&0.14\\
\hline
\hline
    \end{tabular*}

\end{table}

\subsection{Final Photometry}
\label{sub:final_phot}
We measure the final photometry on the host-subtracted image from the previous section in all filters and epochs. The final photometry measured here is reported in Table \ref{tab:im_mags}, and the uncertainties correspond to $\sigma_{\rm{tot}} = \sqrt{\sigma_{\rm{stat}}^2+\sigma_{\rm{plant}}^2+\sigma_{\rm{PSF}}^2}$ where $\sigma_{\rm{stat}}$ is the statistical uncertainty derived from the posterior of each PSF fitting routine, $\sigma_{\rm{plant}}$ is measured from injected source recovery in Section \ref{sub:fakes}, and $\sigma_{\rm{PSF}}$ is the estimated additional uncertainty from performing PSF photometry on level 3 data (Section \ref{sub:psf_drz}). A final source of photometric uncertainty is a systematic uncertainty on the zero-points, which is $\lesssim0.01$mag for all filters and is therefore sub-dominant to the uncertainties derived here (M. Boyer private communication; Boyer et al., in preparation).
\begin{table*}
    \centering
    \caption{\label{tab:im_mags} Photometry measured for each image of SN H0pe in each filter and for each observing epoch.}
    
    \begin{tabular*}{\linewidth}{@{\extracolsep{\stretch{1}}}*{7}{c}}
\toprule
MJD&\multicolumn{1}{c}{Filter}&\multicolumn{1}{c}{Exp. Time (s)}&\multicolumn{1}{c}{$m_{2\rm{a}}$}&
\multicolumn{1}{c}{$m_{2\rm{b}}$}&\multicolumn{1}{c}{$m_{2\rm{c}}$}&\\
\hline
60033&F090W&2490&30.55$\pm$0.47&25.53$\pm$0.05&28.71$\pm$0.11\\
60033&F115W&2490&28.35$\pm$0.11&24.69$\pm$0.06&26.16$\pm$0.07\\
60033&F150W&1889&26.50$\pm$0.08&23.93$\pm$0.07&24.95$\pm$0.07\\
60033&F200W&2104&26.45$\pm$0.08&24.21$\pm$0.08&25.25$\pm$0.08\\
60033&F277W&2104&25.72$\pm$0.14&24.93$\pm$0.14&25.58$\pm$0.14\\
\hline
60057&F090W&1245&$>29.71$&26.44$\pm$0.07&30.23$\pm$0.74\\
60057&F150W&858&27.06$\pm$0.10&24.03$\pm$0.07&25.45$\pm$0.08\\
60057&F200W&1760&26.82$\pm$0.09&24.35$\pm$0.08&25.06$\pm$0.08\\
60057&F277W&1760&26.07$\pm$0.15&25.20$\pm$0.14&25.37$\pm$0.14\\
\hline
60074&F090W&1417&$>29.48$&27.21$\pm$0.09&30.45$\pm$0.81\\
60074&F150W&1245&27.37$\pm$0.12&24.42$\pm$0.07&25.94$\pm$0.08\\
60074&F200W&1760&26.74$\pm$0.09&24.81$\pm$0.08&25.24$\pm$0.08\\
60074&F277W&1760&26.28$\pm$0.15&25.21$\pm$0.14&25.18$\pm$0.14\\
\hline
\hline
    \end{tabular*}
\begin{flushleft}
\tablecomments{Columns are: Modified Julian date, NIRCam filter, exposure time in seconds, and photometry plus final uncertainty for SN image ``$m_{2\rm{a}}$'', ``$m_{2\rm{b}}$'', and ``$m_{2\rm{c}}$'' in AB magnitudes.  Upper limits are 3$\sigma$.}

\end{flushleft}
\end{table*}

\section{Light Curve Fitting}
\label{sec:fitting}
We fit the photometry of the multiple images simultaneously using the SN Time Delays (\sntd) software package \citep{pierel_turning_2019}. Milky Way dust extinction ($E(B-V)=0.019$,  $R_V=3.1$) was based on the maps of \citet{schlafly_measuring_2011} and extinction curve of \citet{fitzpatrick_correcting_1999}. We used a phase-extended version of the BayeSN SN\,Ia SED model \citep{mandel_hierarchical_2022,grayling_scalable_2024}, which covers the full plausible rest-frame phase range of \lensedsn, to fit the measured photometry. The version of BayeSN we used is a variant of the one presented by \citet{ward_relative_2022} and uses the training set detailed therein. The model is trained to cover rest-frame phases as late as $50$ days with linear extrapolation in the space of the logarithm of the SED beyond that. We included $u$-band data in the training, enabling coverage of the F090W photometry of \lensedsn. This is the only model for SN\,Ia light curve evolution with the necessary flexibility and sufficient wavelength and temporal coverage for this dataset (particularly for image 2a), making it difficult to estimate systematic uncertainties associated with a single choice of SN model. We therefore produced a suite of simulations (described in Section \ref{sec:sims}) to quantify the final time-delay uncertainties. In addition to the base template and Milky Way dust noted above, BayeSN includes a rest-frame host-galaxy dust component (parameterized by the $V$-band extinction $A_V$ and ratio of total to selective extinction $R_V$) and a light curve shape parameter $\theta$. (This has a strong negative correlation with SALT2's stretch parameter, $x_1$, meaning positive $\theta$ implies a faster decline rate; see e.g., Figure 4 of \citealp{mandel_hierarchical_2022}.) The free parameters in the model and their status in the fitting process are summarized by Table \ref{tab:fitting}. 


\begin{figure*}
    \centering
    \includegraphics[trim={1cm 1cm 0cm 0cm},clip,width=.65\textwidth]{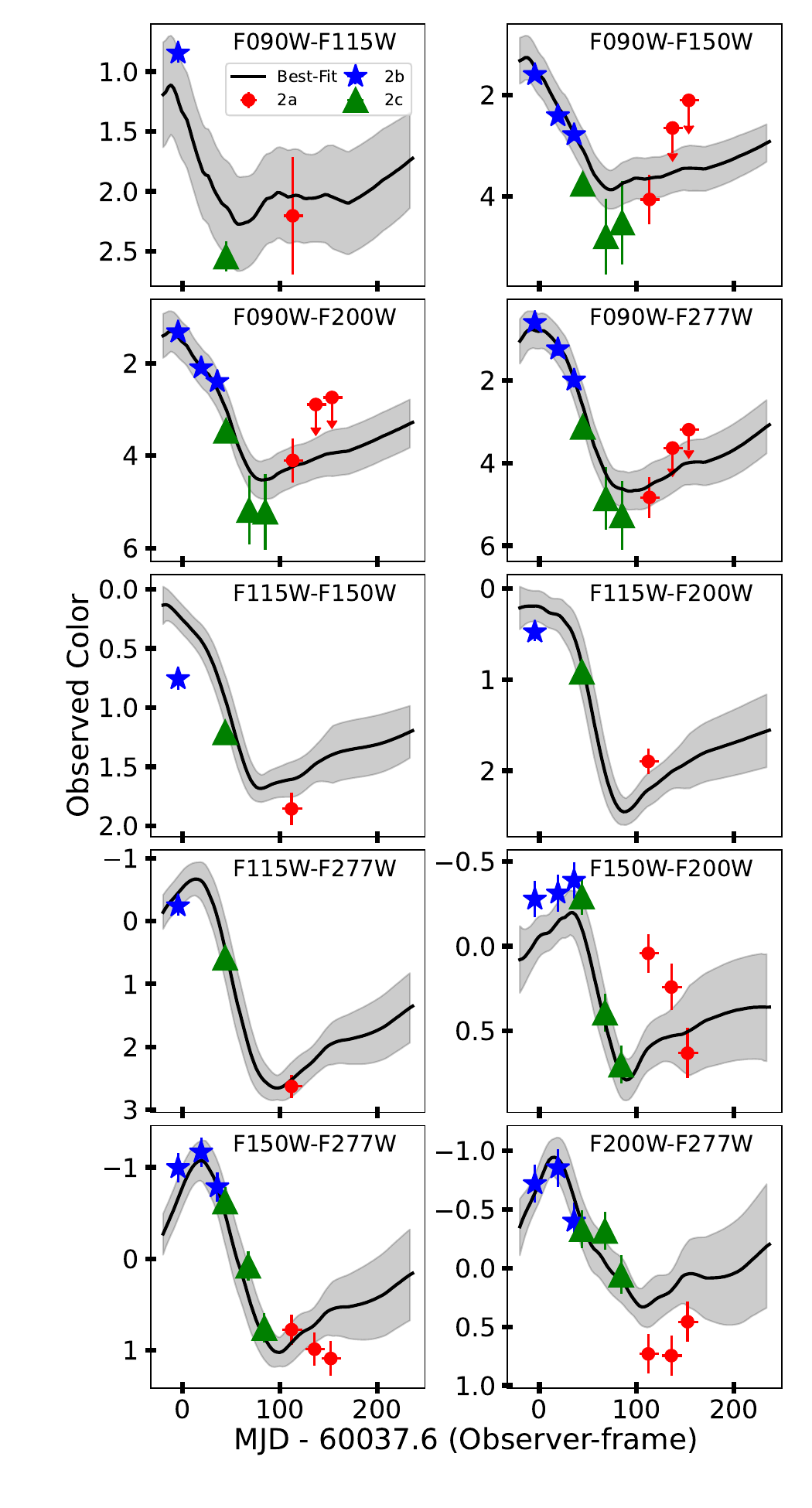}
    \caption{Reconstructed color curves for \lensedsn, after applying the best-fit time delays. The vertical error bars are the photometric precision, while the horizontal error bars are the $16^{\rm{th}}$ and $84^{\rm{th}}$ percentiles simulation recovery time-delay posterior for images 2a and 2c (see Table \ref{tab:sim_results}), and $t_{pk}$ for image 2b. The grey shaded region is uncertainty on the best-fit model from the \sntd Color method (black solid).}
    \label{fig:intrinsic_colorcurve}
\end{figure*}
\begin{figure*}
    \centering
    \includegraphics[width=\linewidth]{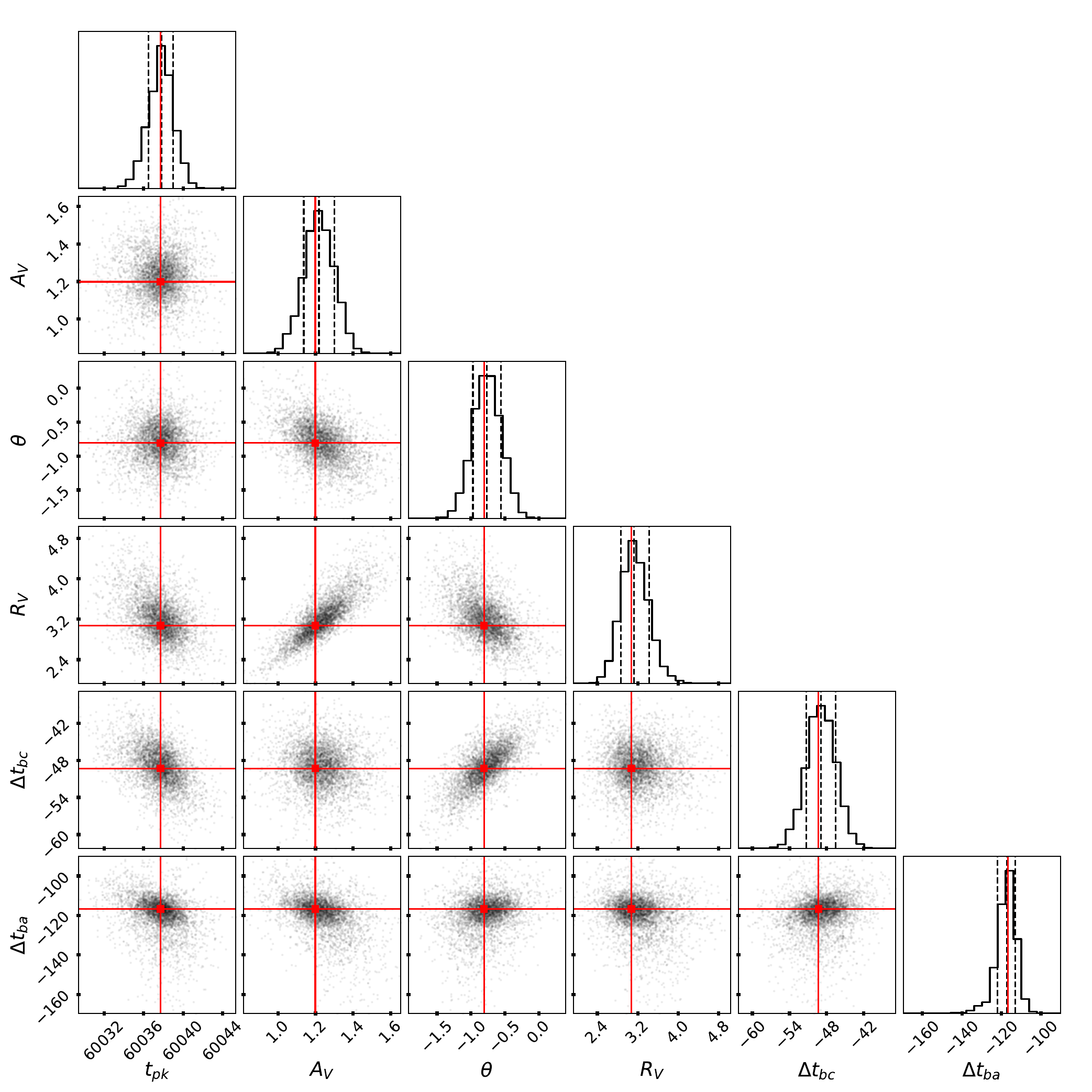}
    \caption{Posterior distributions of the \sntd Color-method fitting of \hst photometry. The dashed vertical lines correspond to the distribution $16^{\rm{th}} , \ 50^{\rm{th}}, \ \rm{and} \ 84^{\rm{th}}$ percentiles.  The reported measurement values (solid red lines) are the maximum-likelihood estimates from these distributions.}
    \label{fig:sntd_corner}
\end{figure*}
\begin{figure}
    \centering
    \includegraphics[trim={.7cm 0cm 2cm 1cm},clip,width=\linewidth]{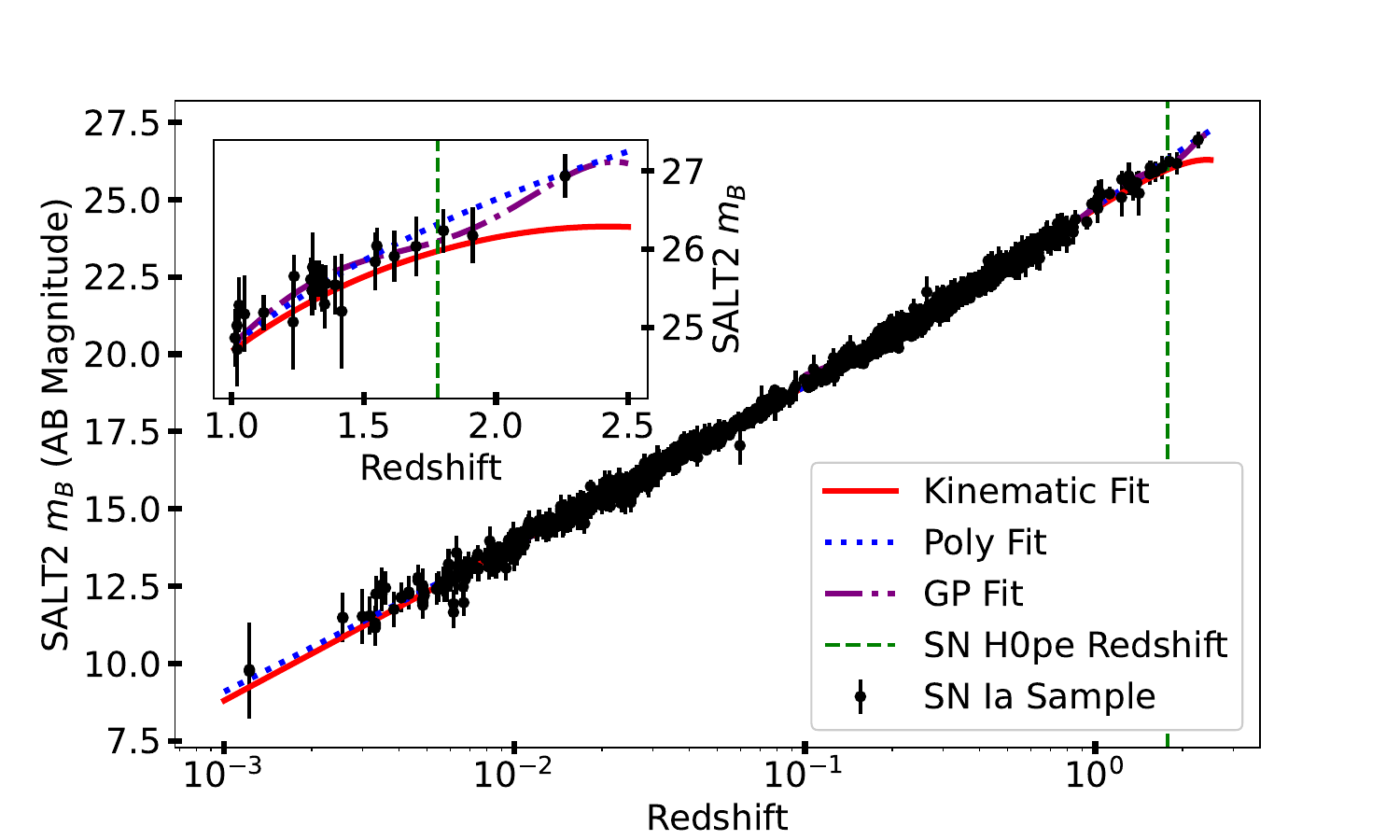}
    \caption{A visual representation of finding the predicted value of $m_B$ for a SN\,Ia at z=\snz, using the SNe\,Ia from \citet{scolnic_complete_2018,brout_pantheon_2022} (black). The data are fit with a second degree polynomial (blue dotted), a gaussian process (purple dash-dot), and a kinematic expansion (red solid). The redshift of \lensedsn is shown as a dashed green line.}
    \label{fig:pantheon_mB}
\end{figure}

\begin{figure*}
    \centering
    \includegraphics[trim={1cm 0cm 2.5cm 1cm},clip,width=\textwidth]{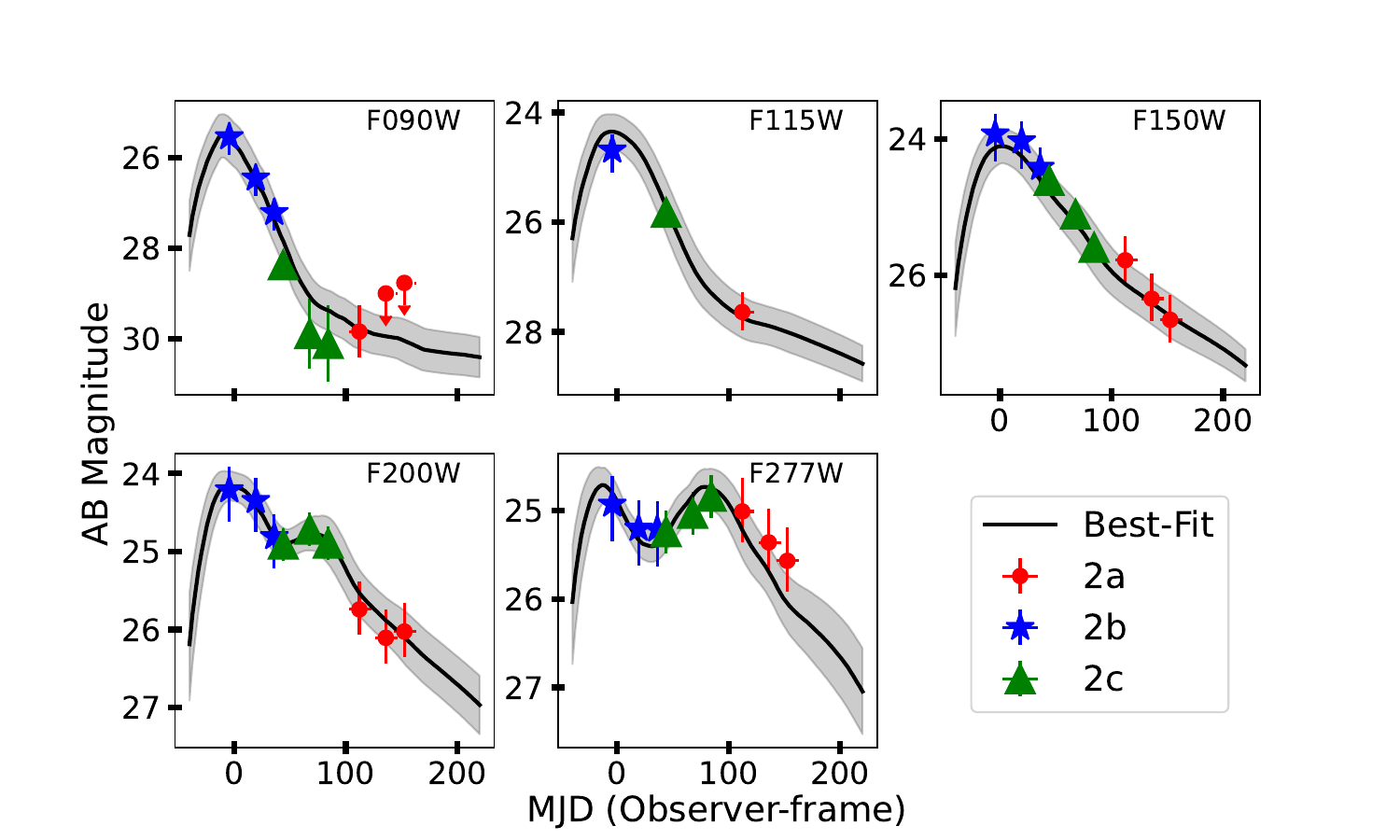}
    \caption{Reconstructed light curve for \lensedsn after applying the best-fit time delays and magnifications relative to image 2b. The vertical error bars are the photometric precision based on the work in Section \ref{sec:photometry} combined with an additional magnification uncertainty from Section \ref{sec:sims}, while the horizontal error bars are the $16^{\rm{th}}$ and $84^{\rm{th}}$ percentiles for the time-delay posteriors from Figure \ref{fig:sntd_corner}. The grey shaded region is the best-fit model from the \sntd Color method scaled to image 2b.}
    \label{fig:intrinsic_lightcurve}
\end{figure*}

\begin{table*}[t!]
    \centering
    \caption{\label{tab:fitting}Summary of the BayeSN parameters used in fitting \lensedsn.}
    
    \begin{tabular*}{\linewidth}{@{\extracolsep{\stretch{1}}}*{5}{c}}
\toprule
Parameter&Description&Free/Fixed&Bounds&Value\\
\hline
$z$&Redshift&Fixed&--&\snz\\
$t_{pk}$ (MJD)&Image 2b time of maximum&Free&$[-20,20]+60033$&$60037.77_{-2.01}^{+2.12}$\\
$\theta$&Light curve shape&Free&[$-2,2$]&$-0.74_{-0.27}^{+0.30}$\\
$A_V$&$V$-band dust extinction&Free&[$0,3$]&$1.21_{-0.11}^{+0.11}$\\
$R_V$&$A_V/E(B-V)$&Free&[$1,6$]&$3.10^{+0.40}_{-0.34}$\\
\hline
\hline
    \end{tabular*}

\end{table*}
\subsection{Time Delays}
\label{sub:time_delays}
With the knowledge that \lensedsn is a SN\,Ia, we are able to leverage the BayeSN SED model \citep{mandel_hierarchical_2022} to measure time delays with the \sntd Color method \citep{pierel_turning_2019,pierel_lenswatch_2023}. This method finds common values for the BayeSN light curve parameters (i.e., shape $\theta$ and dust extinction parameters $R_V$ and $A_V$) amongst all \lensedsn images, while measuring time delays relative to the value of $t_{pk}$ corresponding to image 2b (chosen as the brightest image and last to arrive,  Figure \ref{fig:color_im}). The Color method directly fits using color curves i.e., observed SN colors over time), which removes the need for normalization and relative magnifications. The primary advantage of the Color method is that it is insensitive to achromatic microlensing as any such effect is identically present in all bands (The impacts of chromatic microlensing are discussed in Section \ref{sub:sim_sed_micro}). Degeneracies between magnification ratios and time delays are also removed, but the cost of color curves is fewer clear inflection points compared to light curves in certain filter combinations and phases. The \lensedsn light curve has sufficient features in the color curves to use this method, which optimizes a joint likelihood function including both the SN\,Ia light curve model parameters and the time delays with bounds of $\pm200$ observer-frame days. Figure \ref{fig:intrinsic_colorcurve} shows the measured photometry and model with best-fit parameters and time delays applied, and the results are summarized by Tables \ref{tab:fitting} and \ref{tab:td_mu}. Note that the measured value for $\theta$, $-0.74$, is equivalent to a normal decline rate of $\Delta m_{15}(B)=1.13$ \citep{phillips_absolute_1993}. The measured value of $\Delta t$ in Table \ref{tab:td_mu} includes corrections based on the systematic uncertainties derived in Section \ref{sec:sims} following the methods of \citet{kelly_magnificent_2023}. While some of the points appear discrepant from the model at $\sim1$-$2\sigma$ (e.g., the first epoch for image 2a in F150W-F200W), this has been propagated through to the final time-delay and magnification uncertainties by way of our simulations in Section \ref{sec:sims}, and further improvements will require improved photometry from difference imaging and an improved late-time SN\,Ia model (see Section \ref{sec:conclusion}). The posterior distributions from the fitting are shown in Figure \ref{fig:sntd_corner}. The distributions display an expected covariance between $(R_V,A_V)$, and some covariances between $(\theta,\Delta t_{bc})$ and $t_{pk}$ and the time-delays. The covariances with $t_{pk}$ are also to be expected, while the $(\theta,\Delta t_{bc})$ covariance is likely because the second infrared maximum constrains $\Delta t_{bc}$ and is directly related to $\theta$. While we cannot fit all three images of \lensedsn with existing models other than BayeSN, for consistency we do measure time delays for images 2b and 2c with the SALT3-NIR model. We find agreement within $1\sigma$, and proceed with our BayeSN result as it can applied to all three images.

    


\subsection{Magnifications}
\label{sub:magnifications}
Relative magnifications between the multiple images have been used as a model-independent means of weighting lens models while inferring $H_0$ \citep[i.e.,][]{kelly_constraints_2023}, but absolute magnifications provide a more stringent constraint when measurable. Given their standardizability, SNe\,Ia provide a rare opportunity to measure an absolute magnification, increasing their utility for time delay cosmology \citep[e.g.,][]{birrer_hubble_2022}. As mentioned in the previous section, the \sntd Color method does not measure an overall normalization. We therefore fixed the intrinsic model parameters and time delays to those measured in the previous section and then fit for the image 2b normalization and magnification ratios of images 2a, 2c relative to image 2b. We converted the magnification ratios to absolute magnifications by assuming \lensedsn is standardizable in the rest-frame NIR, where the impacts from dust are mitigated relative to the rest-frame optical. Specifically, we used the peak F277W magnitude ($\sim$rest-frame $Y$-band) measured here from BayeSN including the final measurement uncertainty, which previous work has shown to be an excellent estimator for luminosity distance while fitting for shape and color for normal SNe\,Ia such as \lensedsn \citep[e.g.,][]{pierel_salt3-nir_2022}. We fit the complete sample of spectroscopic SN\,Ia peak $m_B$ measurements \citep{scolnic_complete_2018,brout_pantheon_2022} with three separate methods to obtain the predicted value of $m_B$ at $z=\snz$ (Figure \ref{fig:pantheon_mB}). The three methods are a simple second-order polynomial, a gaussian process, and a kinematic expansion \citep[][; Equation 4 with $j_0=1$]{riess_comprehensive_2022}. We take the average of these methods ($26.13$ AB mag), and propagate the standard deviation to the magnification uncertainties ($0.14$ mag). We then scaled the best-fit BayeSN model to this $B$-band magnitude and predict a non-lensed F277W AB magnitude of $26.98$. We compared the best-fit model-predicted F277W to this reference magnitude for each image to give absolute magnification measurements. This would be fully model-independent given a set of distance measurements using F277W at this redshift, but this is not possible at the current time. However, as the $m_B$ measurements we used as a reference are independent of cosmology, our magnification measurements are still independent of $H_0$. We combined the uncertainties on each measured magnification (Section \ref{sec:sims}) with systematic uncertainties based on the intrinsic scatter of SN\,Ia absolute magnitudes measured by \citet[][0.1\,mag]{scolnic_complete_2018}. They further report zero-point uncertainties of $<0.01 mag$, and we therefore ignore this additional systematic. The measured time delays and magnifications are shown in Table \ref{tab:td_mu}, and the final reconstructed light curve is shown in Figure \ref{fig:intrinsic_lightcurve} where all observations are consistent with the model within $1\sigma$. Once again, we attempt to measure magnifications in the same manner using SALT3-NIR for images 2b and 2c, and find agreement within $1$-$2\sigma$. While this is in slight tension, without more SNe\,Ia at high-$z$ this method must remain model-dependent, and we therefore proceed with the BayeSN result as it can be applied to all three SN images.

\begin{table}
    \centering
    \caption{\label{tab:td_mu}Measured time delays and magnifications}
    
    \begin{tabular*}{\linewidth}{@{\extracolsep{\stretch{1}}}*{3}{c}}
\toprule
Image&$\Delta t$ $(t_i-t_b)$&$\mu$\\
&(Days)&\\
\hline
2a&\dtab&\mua\\
2b&--&\mub\\
2c&\dtcb&\muc\\
\hline
\hline
    \end{tabular*}

\end{table}
\begin{table*}
    \centering
    \caption{ \label{tab:sim_results}The variation in measured time delay and magnification of each simulation scenario from Section \ref{sec:sims}, relative to the measurements of Section \ref{sec:fitting}. }
    
    \begin{tabular*}{\linewidth}{@{\extracolsep{\stretch{1}}}*{7}{c}}
\toprule
Sim Type (Section)&\multicolumn{2}{c}{Image A}&\multicolumn{2}{c}{Image B}&\multicolumn{2}{c}{Image C}\\
&$\delta t$ (days)&$\delta\mu$&$\delta t$ (days)&$\delta\mu$&$\delta t$ (days)&$\delta\mu$\\
\hline
Photometric (\ref{sub:sim_phot})&$0.55_{-7.13}^{+8.74}$&$-0.36_{-1.36}^{+1.07}$&--&$-0.26_{-0.86}^{+1.05}$&$-0.57_{-3.94}^{+2.76}$&$-0.22_{-0.70}^{+0.79}$\\
SED (\ref{sub:sim_sed_base})&$-1.24_{-4.95}^{+4.80}$&$0.08_{-1.61}^{+0.44}$&--&$-0.47_{-0.59}^{+1.26}$&$-0.45_{-2.11}^{+1.91}$&$-0.01_{-1.36}^{+0.70}$\\
Microlensing (\ref{sub:sim_sed_micro})&$-0.20_{-0.82}^{+0.95}$&$0.22_{-0.20}^{+0.19}$&--&$-0.63_{-1.49}^{+2.51}$&$-0.04_{-0.16}^{+0.15}$&$0.30_{-0.47}^{+0.20}$\\
\hline
Combined (\ref{sub:sim_combined})&$-0.97_{-9.29}^{+10.83}$&$0.03_{-1.61}^{+1.38}$&--&$-0.32_{-2.21}^{+3.36}$&$-0.67_{-3.97}^{+3.58}$&$0.28_{-1.07}^{+0.95}$\\
Millilensing (\ref{sub:sim_milli})&--&$0.06_{-0.28}^{+0.43}$&--&$-0.12_{-0.65}^{+0.52}$&--&$0.12_{-0.43}^{+0.74}$\\

Intrinsic SN\,Ia Scatter (\ref{sub:magnifications})&--&$0.00^{+0.40}_{-0.40}$&--&$0.00^{+0.76}_{-0.76}$&--&$0.00^{+0.59}_{-0.59}$\\
$m_B$ Interpolation (\ref{sub:magnifications})&--&$0.00^{+0.56}_{-0.56}$&--&$0.00^{+0.99}_{-0.83}$&--&$0.00^{+0.83}_{-0.83}$\\
\hline
\textbf{Final Uncertainty}&$\mathbf{-0.97_{-9.29}^{+10.83}}$&$\mathbf{0.09_{-1.77}^{+1.60}}$&\textbf{--}&$\mathbf{-0.44_{-2.61}^{+3.61}}$&$\mathbf{-0.67_{-3.97}^{+3.58}}$&$\mathbf{0.40_{-1.54}^{+1.57}}$\\
    \end{tabular*}

\end{table*}



\section{Simulations}
\label{sec:sims}
We simulated variations of the measured color curves both to test the robustness of our fitting procedure and to estimate the statistical and systematic uncertainties of the time-delay and magnification measurements. We created two distinct sets of simulations (photometric variability, Section \ref{sub:sim_phot}; SED variability, Section \ref{sub:sim_sed}) that test different aspects of the fitting uncertainties, described below. We then varied the second set further by introducing the effects of chromatic microlensing (Section \ref{sub:sim_sed_micro}). These simulations serve to illuminate the relative sources of uncertainty for this analysis, and then we create a single final set of simulations that include all of the effects discussed in sections \ref{sub:sim_phot}-\ref{sub:sim_sed_micro} that define our combined uncertainties (Section \ref{sub:sim_combined}). These are added in quadrature with three external (not related to the SED or light curve) sources of systematic uncertainty, namely millilensing (Section \ref{sub:sim_milli}), the intrinsic scatter of SNe\,Ia, and the SN\,Ia $m_B$ population interpolation uncertainty (both discussed in Section \ref{sub:magnifications}). The results of this section are summarized by Table \ref{tab:sim_results}.

\subsection{Random Photometric Realizations}
\label{sub:sim_phot}
We created mock observations drawn from a normal distribution centered at each measured flux with  $\sigma$ corresponding to the measured photometric uncertainty (Table \ref{tab:im_mags}). We created simulated color curves and generated 100 realizations of each SN image. We ran the identical \sntd Color method and magnification measurement used in the previous section on each of the realizations, and  Table \ref{tab:sim_results} shows the results. Due to the relatively large systematic uncertainties in the photometry (Table \ref{tab:plant_sigmas}), varying the photometry has quite a large impact on the time-delay measurements. The effect is most significant for Image 2a, where the bulk of the time-delay inference comes from the F277W photometry, which has the largest uncertainties. The template imaging planned for \lensedsn in \textit{JWST} Cycle 3 will both improve the photometry in F277W and add the F356W/F444W filters, likely reducing this uncertainty significantly. 

\subsection{SED-Based Simulations}
\label{sub:sim_sed}
\subsubsection{Baseline simulations}
\label{sub:sim_sed_base}
Mock color curves were also created using the BayeSN model to explore model and/or wavelength dependent systematics. These simulations include realizations of residual intrinsic light-curve variation (beyond the principal effect of light-curve shape $\theta$) drawn from BayeSN's learned distribution of intrinsic residuals ($\epsilon$; see \citealp{mandel_hierarchical_2022} \S5.2.2 for details). In this way, our simulations should capture the intrinsic diversity of SN Ia color curves. The simulations include a range of values for the BayeSN light curve shape ($\theta$) and dust extinction ($A_V$) values based on independent constraints from the NIRSpec data (Chen et al., 2024). The spectra show \lensedsn to be a normal SN\,Ia, and therefore we simulated SNe\,Ia over the reasonable population values for $\theta$ and $A_V$ of $-2\leq\theta\leq2$ and $0\leq A_V\leq1.5$ given $R_V=3.1$ \citep{mandel_hierarchical_2022}. This is the best-fit value of $R_V$ from our light curve fitting, and we fix it here as it is highly covariant with $A_V$ and has little impact on measured time-delays or magnifications (Figure \ref{fig:sntd_corner}).  We then simulated $100$ color curves with cadence and photometric precision matching those described above but with random offsets to the relative time delays and magnifications between images. Again the results of running the identical \sntd Color method and magnification measurement used in Section \ref{sec:fitting} on each of the realizations are summarized in Table \ref{tab:sim_results}. The results from fitting this set of simulations, which essentially includes both the model uncertainty and intrinsic variability of SNe\,Ia as a function of wavelength, is the largest source of uncertainty for both time delays and magnifications. The most promising means of improving this uncertainty would be to add the F356W and F444W filters, which provide additional light curve information particularly at late times, and to continue improving SN\,Ia SED models in the NIR and at late times (beyond 50 rest-frame days after peak brightness).

\subsubsection{The impacts of microlensing}
\label{sub:sim_sed_micro}

Microlensing from small perturbers such as stars in the lens plane is well known to potentially impact time-delay measurements \citep[e.g.,][]{dobler_microlensing_2006,goldstein_precise_2018,pierel_turning_2019,huber_strongly_2019,huber_holismokes_2022}. The effects of microlensing are not generally achromatic because the specific intensity profiles for SNe vary for different filters. Therefore the expanding SN shell interacts with the microlensing caustics, and the resulting magnification is chromatic \citep{goldstein_precise_2018,foxley-marrable_impact_2018,huber_strongly_2019}. For SNe\,Ia, microlensing is essentially achromatic in the first $\sim$3 rest-frame weeks after explosion \citep[i.e., the ``achromatic phase,''][]{goldstein_precise_2018,huber_holismokes_2021}. The method of measuring time delays with color curves is therefore valuable because it removes the uncertainty of macrolensing and achromatic microlensing from the fit. The observations of (at least) image 2b of \lensedsn are likely in the achromatic phase, but it is plausible that the remaining observations are impacted by chromatic microlensing.


\begin{table}[h!]
    \centering
    \caption{\label{tab:micro_params}Parameters used to create microlensing magnification maps for each SN image.}
    
    \begin{tabular*}{\linewidth}{@{\extracolsep{\stretch{1}}}*{3}{c}}
\toprule
Image&$\kappa$&$\gamma$\\
\hline
2a& $0.563$&$0.193$\\
2b&$0.699$&$0.445$\\
2c&$0.576$&$0.246$\\
\hline
\hline
    \end{tabular*}

\end{table}
We produced a magnification map for each SN image and convolved the SN\,Ia light profiles from four theoretical models \citep{suyu_holismokes_2020,huber_holismokes_2021} with each magnification map ($1000$ random positions per model for a total of $4\,000$ convolutions per map)  in the manner of \citet{huber_strongly_2019}. The parameters used to generate the magnification maps are the lens surface-mass density scaled to the critical value ($\kappa$), the image shear ($\gamma$), and the smooth matter fraction ($s$). Here $\kappa$ and $\gamma$ at each SN image position are the median of six independent lens model predictions from \citet{pascale_h0_2024}, and we assume a value of $0.99$ for $s$ across all maps, consistent with \citet{kelly_magnificent_2023}. Table \ref{tab:micro_params} shows the $\kappa$ and $\gamma$ values used for each SN image. We selected $40$ random chromatic microlensing curves for baseline simulation from the previous section and varied the flux in each band by the phase- and wavelength-dependent magnification curves. Our total number of microlensed simulations was therefore $4\,000$. 
Finally, we ran the \sntd Color method and magnification measurement used in Section \ref{sec:fitting} one more time on each of the realizations.  Table \ref{tab:sim_results} summarizes the results. 

\begin{figure}
    \centering
    \includegraphics[trim={1.6cm, .6cm 0cm 1cm},clip,width=.5\textwidth]{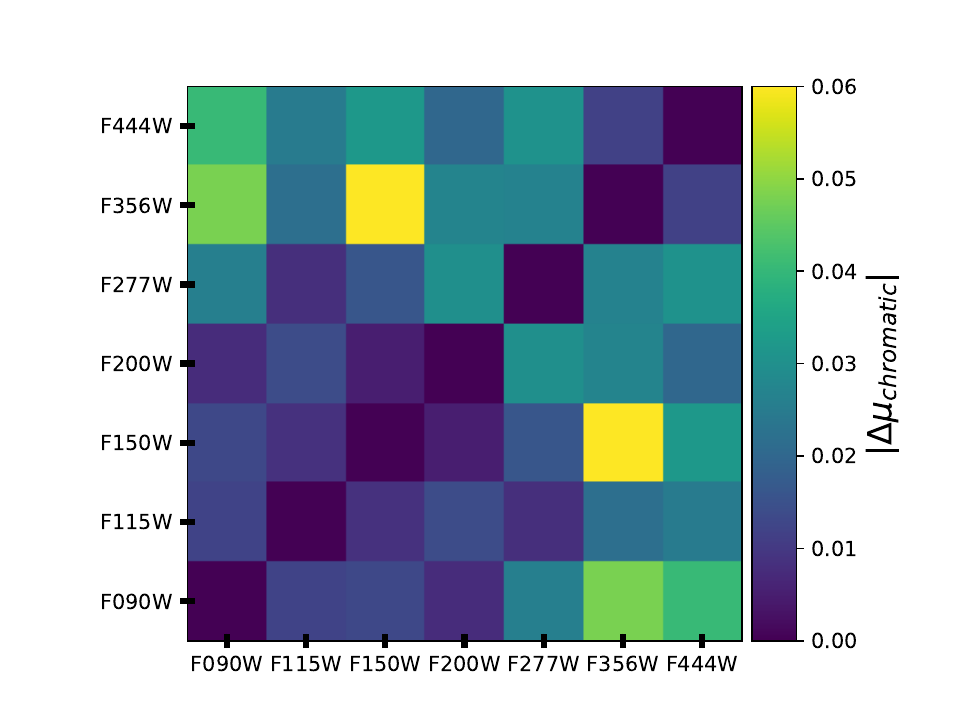}
    \caption{The level of chromatic microlensing expected from our simulations in the case of \lensedsn (across all three SN images) showing all the possible color combinations. The color bar corresponds to the shift in magnitudes. For example, the cell associated with F150W and F200W corresponds to the F150W$-$F200W color.}
    \label{fig:chromatic}
\end{figure}

For completeness, we checked for the level of chromatic microlensing effects that could be impacting our observed photometry by examining the maximum chromatic deviation due to microlensing over the course of the full light curve for all filter combinations. The results are summarized in Figure \ref{fig:chromatic}, which suggests a maximum of $\sim$0.06 mag of chromatic microlensing. The impact overall is relatively low, reaching a maximum of $\gtrsim$0.05 mag in the F356W and F444W filters, and is generally $<$0.02 mag in the filters used in this analysis. This is consistent with the small additional uncertainty due to microlensing found by fitting the simulated light curves. If a future time-delay analysis includes the F356W and F444W filters, particular care should be taken with respect to chromatic microlensing, but the impacts are small for the present analysis. This is consistent with the results of fitting the simulations, which found the extra uncertainty due to microlensing to be quite small compared to the other sources of uncertainty explored here. 

\subsection{Combined Simulation}
\label{sub:sim_combined}
We produce one final set of $1000$ simulated versions of \lensedsn, which include all the sources of variability discussed above, ensuring that the full covariance of these uncertainties is correctly included. Our final uncertainties are the result of fitting these simulations, which contain all sources of uncertainty we are able to simulate. We combine this with an additional systematic uncertainty from millilensing, discussed in the following section, along with systematics of $0.1$mag and $0.03$mag due to the intrinsic scatter of SNe\,Ia and the uncertainty in our $m_B$ interpolation (see Section \ref{sub:magnifications}), with the results shown in Table \ref{tab:sim_results}. The final row in Table \ref{tab:sim_results} includes systematic shifts, which were used to correct each actual measurement from sections \ref{sub:time_delays} and \ref{sub:magnifications}. 

\subsection{The Impacts of Millilensing}
\label{sub:sim_milli}
\citet{kelly_magnificent_2023} found that millilensing, which is additional lensing caused by dark matter halos (referring to subhalos associated with the cluster and halos along the line of sight), caused an additional $\sim$10\% uncertainty in the measured magnifications for SN Refsdal. The impact of millilensing is achromatic and therefore will not impact our time delay measurements but could plausibly impact the image magnifications at a similar level. Using the ray-tracing technique presented by \citet{gilman_probing_2019}, we computed the expected millilensing due to dark-matter cluster subhalos and line-of-sight halos for a projected mass density in substructure
bracketing the range of expected values based on theoretical and observational arguments \citep{gilman_warm_2020}.  This corresponds to a $1$--$10\%$ fraction of mass in dark matter subhalos with masses in the range $10^6\,M_\odot$ to $10^{10}\,M_\odot$. Halos more massive than $10^{10}\,M_{\odot}$ would likely contain a
visible counterpart, and halos less massive than $10^{6}\,M_{\odot}$ are too small to significantly alter the image magnifications. We modeled
the line-of-sight structure using a Sheth–-Tormen halo mass
function \citep{sheth_ellipsoidal_2001}, placed uniformly (in 2D) around each SN image and along a cylindrical line-of-sight volume with diameter equal to the comoving distance to the cluster times $1\arcsec$. We report the additional uncertainty in magnification due to millilensing in Table \ref{tab:sim_results}, where we assume the median scenario of a $5\%$ dark matter contribution from subhalos, finding that magnification due to millilensing ranges
from a $\sim$4\% to $\sim$8\% effect among the three SN images.

\section{Conclusions}
\label{sec:conclusion}

\lensedsn is the first multiply-imaged SN\,Ia and second  multiply-imaged SN capable of delivering a competitive $H_0$ measurement. 
Time delays (relative to the last image to arrive, 2b) are $\Delta t_{ab}=$\dtab, $\Delta t_{cb}=$\dtcb observer-frame days.
Leveraging the fact that \lensedsn has a standardizable light curve, 
absolute magnification estimates are $\mu_a=$\mua, $\mu_b=$\mub, $\mu_c=$\muc. These time-delay measurements will be included in an upcoming inference of $H_0$ \citep{pascale_h0_2024}. While the inferred value of $H_0$ is not shown here, if we assume a fiducial result of $H_0=70\rm~{km s^{-1}\,Mpc^{-1}}$ then the uncertainty from this time-delay measurement alone would be ${\sim}3.9~\rm{km s^{-1}\,Mpc^{-1}}$, which is then added in quadrature to the lens model uncertainty.

While we are unable to fully explore possible systematic uncertainties introduced by using a specific model to analyze \lensedsn, we attempted to mitigate this by using simulations to estimate the uncertainties, including a model of intrinsic SN\,Ia variability in the generative model to account for possible shortcomings in the model. We modeled the possible magnification uncertainty due to millilensing from dark matter subhalos, which does not impact time delay measurements but contributes an additional magnification uncertainty of $\delta \mu_a=-0.06_{-0.28}^{+0.43}, \ \delta \mu_b=0.12_{-0.65}^{+0.52}, \ \delta \mu_c=-0.12_{-0.43}^{+0.74}$. We also considered the possible impacts of chromatic microlensing, which are relatively small ($\delta t_{ab}=-0.20_{-0.82}^{+0.95}, \ \delta t_{cb}=-0.04_{-0.16}^{+0.15}$ days and $\delta \mu_a=0.22_{-0.20}^{+0.19}, \ \delta \mu_b=-0.63_{-1.49}^{+2.51}, \ \delta \mu_c=0.30_{-0.47}^{+0.20}$) as the SN is well-offset from cluster-member galaxies, and the observations of image 2b are in the achromatic phase of SN\,Ia microlensing variability, which limits the impact for time-delay measurements \citep{goldstein_precise_2018,huber_holismokes_2021}. The exception is the magnification measurement for image 2b, which based on our simulations may be significantly impacted by microlensing, as it has higher values for both $\kappa$ and $\gamma$ relative to images 2a and 2c.

The most significant sources of uncertainty in this analysis are the photometry ($\delta t_{ab}=0.55_{-7.13}^{+8.74}, \ \delta t_{cb}=-0.57_{-3.94}^{+2.76}$ days and $\delta \mu_a=-0.36_{-1.36}^{+1.07}, \ \delta \mu_b=-0.26_{-0.86}^{+1.05}, \ \delta \mu_c=-0.22_{-0.70}^{+0.79}$), due to the brightness of Arc 2 (particularly at longer wavelengths), and uncertainties from the SN\,Ia SED model BayeSN ($\delta t_{ab}=-1.24_{-4.95}^{+4.80}, \ \delta t_{cb}=-0.45_{-2.11}^{1.91}$ days and $\delta \mu_a=0.08_{-1.61}^{+0.44}, \ \delta \mu_b=-0.47_{-0.59}^{+1.26}, \ \delta \mu_c=-0.01_{-1.36}^{+0.70}$). The latter issue is likely to be resolved with time, as more NIR data are collected from (in particular) the \textit{Nancy Grace Roman Space Telescope}. While we implemented a method to remove host galaxy light using a novel lens-modeling technique, large residuals remained at the longest wavelengths, and we were unable to detect the SN despite predicted brightnesses that were above our detection threshold. A set of \textit{JWST}/NIRCam template images in the same filters is planned for Cycle 3 (PID 4744), which will greatly improve the LW channel photometry and thereby the time-delay measurement precision. For now, we used planted PSF models to estimate the photometric precision on our final \lensedsn photometry. 

Upcoming surveys such as the Vera C. Rubin Observatory Legacy Survey of Space and Time \citep[LSST;][]{ivezic_lsst_2019} and the \textit{Nancy Grace Roman Space Telescope} High Latitude Time Domain Survey \citep[HLTDS;][]{pierel_projected_2021,rose_reference_2021} are expected to deliver dozens of useful galaxy-scale multiply-imaged SNe, but the number of expected cluster-lensed SNe similar to \lensedsn and SN Refsdal is uncertain. \textit{Roman} in particular, with its $\sim$2000 deg$^2$ High Latitude Survey (HLS) and $\sim$20 deg$^2$ HLTDS, will observe $\sim$23\,000 massive galaxy clusters \citep{eifler_cosmology_2021}. A subset of these clusters have already been accurately modelled using observations from programs such as the Reionization Lensing Cluster Survey \citep[RELICS;][]{coe_relics_2019}, Hubble Frontier Fields \citep[HFF;][]{lotz_fontier_2017} and \textit{JWST} UNCOVER \citep{weaver_uncover_2023}. This allows an estimate of the multiply-imaged SN yields in the area strongly lensed by the clusters (Bronikowski et al., in preparation). While the \textit{Roman} HLTDS observing fields have not yet been selected, it is important to explore which of the massive clusters provides the optimal prospects for discovering strongly lensed supernovae in order to maximise the potential for a precise measurement of $H_0$, facilitated by the fact that these systems offer long time delays and wide image separations. As a result, cluster-lensed SNe are likely to continue as an extremely valuable subset of the future cosmological sample.


\clearpage

\begin{center}
    \textbf{Acknowledgements}
\end{center}

This paper is based in part on observations with the NASA/ESA Hubble Space Telescope obtained from the Mikulski Archive for Space Telescopes at STScI. We thank the DDT and JWST scheduling team at STScI for extraordinary effort in getting the DDT observations used here scheduled quickly. The specific observations analyzed can be accessed via \dataset[DOI: 10.17909/qgt8-e846]{https://doi.org/0.17909/qgt8-e846}"; support was provided to JDRP and ME through program HST-GO-16264. JDRP is supported by NASA through a Einstein
Fellowship grant No. HF2-51541.001 awarded by the Space
Telescope Science Institute (STScI), which is operated by the
Association of Universities for Research in Astronomy, Inc.,
for NASA, under contract NAS5-26555. BLF is supported by the JWST DDT program 4446. RAW acknowledges support from NASA \textit{JWST} Interdisciplinary Scientist grants NAG5-12460, NNX14AN10G and 80NSSC18K0200 from GSFC. CG, SS, and AA acknowledge financial support through grants PRIN-MIUR 2017WSCC32 and 2020SKSTHZ. The work of LAM was carried out at the Jet Propulsion Laboratory,
California Institute of Technology, under a contract with
NASA. Part of the work by LAM was performed at Aspen Center for Physics, which is supported by National Science Foundation grant PHY-2210452. SD acknowledges support from the Marie Curie Individual Fellowship under grant ID 890695 and a Junior Research Fellowship at Lucy Cavendish College. MG acknowledges support from the European Union’s Horizon 2020 research and innovation programme under ERC Grant Agreement No. 101002652 and Marie Skłodowska-Curie Grant Agreement No. 873089. SB acknowledges support by Stony Brook University. JH acknowledges support from the University of Cambridge’s IoA summer internship. SH and RC thank the Max Planck Society for support through the Max Planck Research Group for SHS. This project has received funding from the European Research Council (ERC) under the European Union’s Horizon 2020 research and innovation programme (LENSNOVA: grant agreement No 771776). AA has received funding from the European Union’s Horizon 2020 research and innovation programme under the Marie Skłodowska-Curie grant agreement No 101024195 — ROSEAU. AZ acknowledges support by Grant No. 2020750 from the United States-Israel Binational Science Foundation (BSF) and Grant No. 2109066 from the United States National Science Foundation (NSF); by the Ministry of Science \& Technology, Israel; and by the Israel Science Foundation Grant No. 864/23. SHS thanks the Max Planck Society for support through the Max Planck Fellowship. This research is supported in part by the Excellence Cluster ORIGINS which is funded by the Deutsche Forschungsgemeinschaft (DFG, German Research Foundation) under Germany's Excellence Strategy -- EXC-2094 -- 390783311. JH was supported by a VILLUM FONDEN Investigator grant (project number 16599). TT acknowledges support from NASA through grant HST-GO-16264. CL acknowledges support from the National Science Foundation Graduate Research Fellowship under grant No. DGE-2233066. FP acknowledges support from the Spanish Ministerio de Ciencia, Innovación y Universidades under grant numbers PID2019-110614GB-C21 and PID2022-141915NB-C21. XH acknowledges the University of San Francisco Faculty Development Fund. CG is supported by a VILLUM FONDEN Young Investigator Grant (project number 25501). TP acknowledges the financial support from the Slovenian Research Agency (grant P1-0031). This work has been enabled by support from the research project grant ‘Understanding the Dynamic Universe’ funded by the Knut and Alice Wallenberg Foundation under Dnr KAW 2018.0067. PLK acknowledges funding from NSF grants AST-1908823 and AST-2308051. S.T. was supported by the European Research Council (ERC) under the European Union’s Horizon 2020 research and innovation programme (grant agreement no.\ 101018897 CosmicExplorer). I.P.-F. acknowledges support from the Spanish State Research Agency 
(AEI) under grant number PID2019-105552RB-C43.

\clearpage

\bibliographystyle{aasjournal}


\end{document}